# Uniaxial strain tuning of functional properties in antiferroelectric $PbZrO_3$ membrane capacitors

*Umair Saeed[1,2,3*‡], Huazhang Zhang[4,5*‡], David Pesquera[1], Fernando Gómez-Ortiz[5], Jessica Padilla- Pantoja[1], Jose Santiso[1], Philippe Ghosez[5], Gustau Catalan[1,6]*

[1]Catalan Institute of Nanoscience and Nanotechnology 5 (ICN2), CSIC and BIST, Campus UAB, Bellaterra, Barcelona, 08193 Catalonia.

[2]Universitat Autònoma de Barcelona, Plaça Cívica, Bellaterra, Barcelona, 08193 Catalonia.

[3]Department of Energy Conversion and Storage, Technical University of Denmark, Anker Engelunds Vej 1, Kongens Lyngby 2800, Denmark.

[4]School of Physics and Mechanics, Wuhan University of Technology, Wuhan 430070, China.

[5]Theoretical Materials Physics, Q-MAT, University of Liège, B-4000 Sart-Tilman, Belgium.

[6]ICREA - Institució Catalana de Recerca i Estudis Avançats, Barcelona, 08010 Catalonia.

* These authors contributed equally to this work.

Corresponding authors: ‡ umasa@dtu.dk

‡ zhanghz782@126.com

**Abstract**

Lead zirconate ($PbZrO_3$, PZO) is a prototypical antiferroelectric (AFE) system promising for applications such as energy storage and neuromorphic computing. However, in conventional thin-film architectures, its functional potential is often limited by substrate clamping. Although emerging freestanding membrane techniques offer an approach to release this clamping and enable dynamic strain engineering, a systematic understanding of how external strain modulates the switching between AFE and ferroelectric (FE) states is missing. In this study, we investigate the uniaxial strain effects on PZO membrane capacitors transferred onto flexible substrates via a gold-bonding method. We demonstrate that in-plane dynamic strain modulates the AFE-FE switching and functional properties. Corroborated by atomistic simulations, we reveal that the strain-dependent evolution arises from the superposition of the intrinsic strain dependence of specific domain orientation variants and extrinsic strain-induced redistribution of the relative population of different domain variants. These findings provide a predictive understanding of uniaxial strain tuning of AFE-FE switching in PZO, and pave the way for exploiting the functional potential of freestanding antiferroelectric membranes.

Lead Zirconate, $PbZrO_3$ (PZO) has long been regarded as the prototypical antiferroelectric (AFE) compound [1-7], serving as a foundational system for understanding and exploiting AFE properties in applications such as high-energy-density storage, electrocaloric cooling, thermal transport modulation, and the emerging neuromorphic computing [8-12]. These functionalities originate from electric field induced switching between the AFE and ferroelectric (FE) states [13-15]. Conventionally, AFE thin films are grown on rigid substrates, and the switching behavior is strongly affected by substrate clamping [16, 17], which constrains lattice deformation and limits functional potential. Recent advances in membrane technologies have demonstrated that releasing films from the rigid substrates can improve structural integrity and enable exceptional mechanical flexibility and elasticity [18-24]. For AFEs, freestanding capacitors show faster switching dynamics and narrower hysteresis without residual polar domains in the absence of voltage, leading to enhanced energy storage density and efficiency compared to clamped films [25].

Beyond simply removing clamping effects, freestanding membranes also provide a unique platform for dynamic strain engineering [23, 24, 26]. Unlike traditional films attached on rigid substrates, where the strain state is fixed upon synthesis and cannot be actively altered, freestanding membranes allow for the controlled application of variable strains. Since the AFE-FE switching is inherently coupled with lattice deformation [27, 28], external mechanical strain is expected to effectively modulate the switching behavior. Despite this potential, a systematic understanding of how strain modulates the AFE-FE switching in PZO membranes remains elusive. Unveiling the strain-property relationship is therefore essential for rationally tuning the switching response and exploiting the mechanical tunability of PZO in flexible electronic applications.

In this study, we investigate the dynamic strain tuning of functional properties in PZO membrane capacitors transferred onto flexible substrates. We demonstrate that uniaxial strain applied through substrate bending effectively tunes the AFE-FE switching behavior in a non-trivial manner: compressive strain favors sharper and narrower double-hysteresis loops, whereas tensile strain favours much broader double-loops that tend to overlap and yield residual polarization at 0 volts. These results are somewhat surprising, as compressive in-plane strain normally tends to favour ferroelectricity, yet we observe the opposite trend. Corroborated by atomistic simulations, we reveal that the evolution of switching hysteresis under uniaxial strain originates from the intrinsic strain dependency of different domain variants, as well as strain-induced reorientation among different domain variants, i.e. the change in their relative populations. These results provide microscopic insight into strain-controlled AFE-FE switching and provide guidelines for the rational design of strain-tunable antiferroelectric devices.

## Experimental results

### *Gold bonding for efficient strain transfer*

Robust adhesion between the membrane and the supporting substrate is essential for efficient strain transfer. However, the conventional method of transferring the membrane involves the membrane adhering to the flexible substrate via Van der Waals forces, which may lead to interfacial sliding and partial strain relaxation under large applied strains. To overcome this limitation, we employ a gold bonding strategy, a technique well-established in the semiconductor industry [29, 30], as schematically illustrated in Fig. 1a. The process began with a $SrRuO_3$ (SRO)/PZO/SRO epitaxial stack grown on $SrTiO_3$ (001) substrate (experimental details provided in supplementary Section S1). We first deposited gold films of 5 nm thickness onto both the epitaxial stack and the flexible Kapton substrate via sputtering, an approach allows for strong bonding between the gold and the underlying materials. We then brought the two gold layers in contact, clamped them together using a house-made clamping fixture to apply pressure, and annealed the clamped system in vacuum at 300 °C for 8 hours. This process facilitates the two gold layers to bond via interdiffusion, firmly attaching the membrane capacitor to the Kapton substrate and enabling a reliable strain transfer. The assembly was then immersed in water to allow the sacrificial layer ($Sr_3Al_2O_6$) to dissolve and remove the substrate. The top SRO was then patterned to form circular electrodes.

Controlled uniaxial strain on the membrane capacitor is applied by bending the Kapton substrate onto a copper stage with defined curvature. By varying the bending direction and curvature radius, both in-plane uniaxial tensile and compressive strains with tunable magnitudes can be applied. We use different radius of curvature protrusions ranging from 5.0 mm to 0.5 mm in the positive curvature case for tensile strain and negative curvature of 3.6 mm radius for compressive strain. The compressive strain is limited to relatively lower values due to geometry restrictions for contacting the probes to the top electrodes.

A potential concern with the substrate bending is the concomitant generation of a strain gradient equal to the curvature, which could introduce flexoelectric effects in the functional layer. This side effect can be effectively minimized by choosing a significantly thick flexible substrate so that high values of strains can be achieved (which are proportional to the distance of the membrane from the neutral plane and therefore to substrate thickness) at sufficiently low curvatures (which is equal to the strain gradients and is independent of substrate thickness). In our case, the thickness of Kapton substrate (120 µm) is nearly three orders of magnitude larger than that of the membrane capacitor (370 nm), positioning the neutral plane 54 µm from the capacitor/Kapton tape interface (Fig. 1b). Quantitative estimation confirms that even under the smallest bending radius/largest curvature (0.5 mm, corresponding to maximum applied curvature of 1766.8 $m^{-1}$), the resulting flexoelectric field is 0.035 kV $cm^{-1}$ and the flexoelectric polarization is only $6.9 \times 10^{-4}$ μC $cm^{-2}$, which are negligibly small (we calculate these values using the known flexoelectric coefficients of $PbZrO_3$ [31] see Supplementary Section S2). These considerations ensure that the observed strain-dependent switching behavior predominantly reflects the response of the PZO membranes to uniform uniaxial strain.

To verify the transfer of strain to the membrane capacitor, we performed X-ray diffraction (XRD) measurements to quantify the lattice deformation. Figure 1c compares the XRD patterns of the flat stack and the stack with a bending radius of 5.0 mm, corresponding to an in-plane uniaxial tensile strain of 1.07%. In both cases, the PZO membrane exhibits coexistence of two domain orientations, which are $(120)_o$ domains and $(001)_o$ domains in the orthorhombic *Pbam* setting, corresponding to the pseudocubic $(100)_{pc}$ and $(001)_{pc}$ orientations, respectively. For the flat membrane, the extracted pseudocubic lattice parameters are close to bulk values: 0.416 nm for the $(120)_o$ domain, and 0.411 nm for the $(001)_o$ domain. Upon bending, all PZO diffraction peaks shift systematically toward higher angles, indicating the development of out-of-plane compressive strain. The calculated out-of-plane strains are −0.40% for the $(120)_o$ domain and −0.38% for the $(001)_o$ domain. Using the measured out-of-plane strain and bulk Poisson's ratio of about 0.34 [32], the expected in-plane strain is 1.11% for PZO, very close to the geometric strain calculated from curvature. This close match confirms that the bending-induced strain is effectively transferred to the PZO membrane.

In addition to the shift of XRD peaks, we also observed variations in the relative peak intensities. Specifically, the intensity of the $(004)_o$ peak relative to the $(240)_o$ peak increases upon bending, indicating an increased fraction of $(001)_o$ domain under tensile strain (see XRD fittings in Supplementary Section S3). The variation in peak intensities indicates a redistribution of domain population, a mechanism that will be further corroborated by the atomistic simulations discussed in Section 3.

XRD on smaller ROC protrusions was not possible due to the XRD spot size being significantly larger, and the comparatively smaller areas of the curved membrane as compared to the flatter regions.

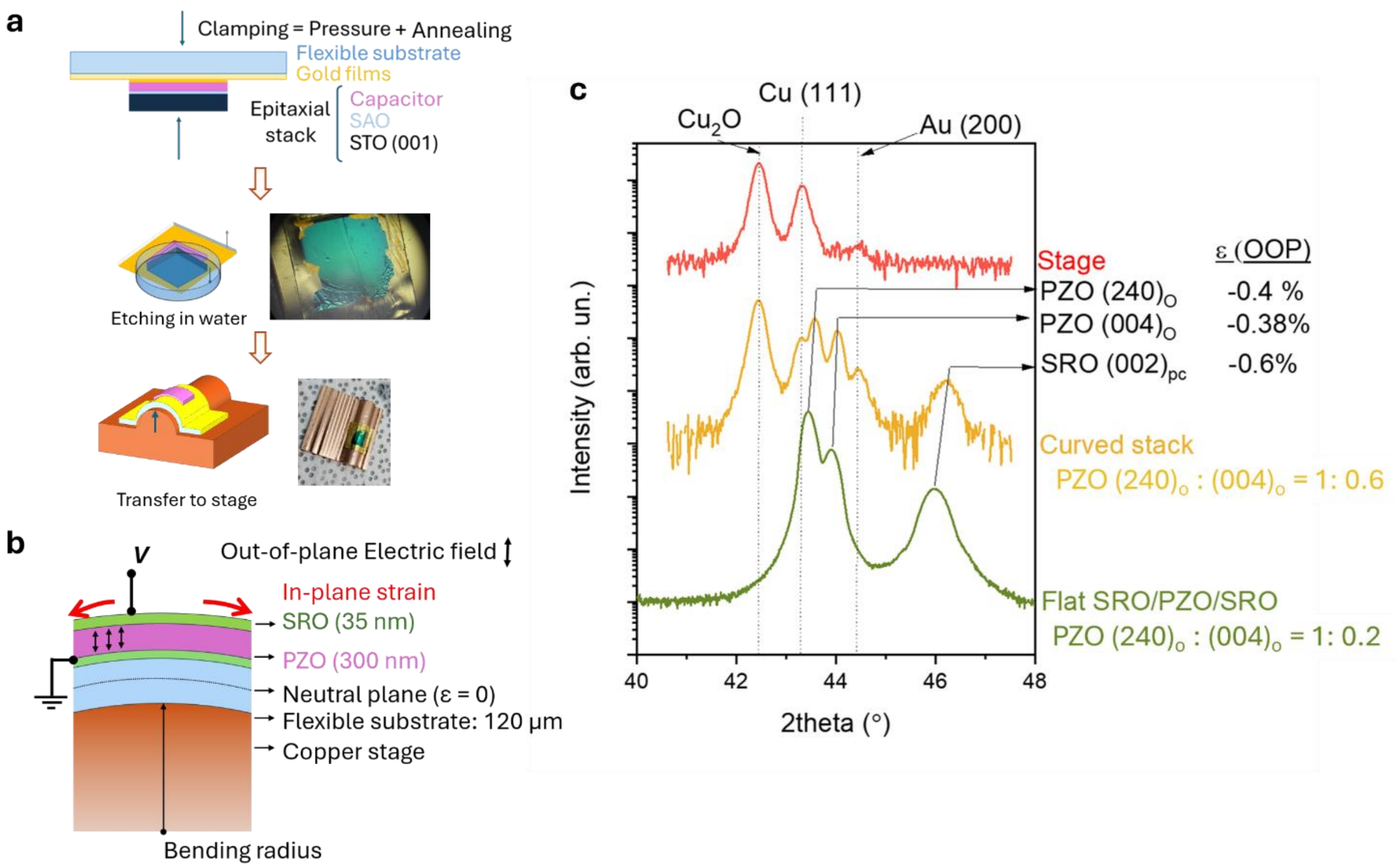

***Figure 1. Gold bonding transfer of PZO freestanding membrane onto flexible Kapton substrates.*** *(a) Schematic illustration and optical images of gold bonding process for transferring large-area membranes onto flexible substrates. (b) Cross-sectional schematic of the sample stack on a copper stage. Electric field is applied in the out-of-plane direction as indicated by double sided black arrows within the PZO layer, while red arrows indicate the primary in-plane strain direction. (c) XRD patterns for the flat membrane capacitor, a capacitor bent to a radius of curvature of 5 mm, and the copper stage reference. The patterns of the bent membrane show a shift toward higher 2θ angles, and an increase in relative intensity of the* $(004)_o$ *peak relative to the* $(240)_o$ *peak.*

*Functional properties*

To investigate the uniaxial strain modulation of functional properties of the PZO membranes, we measured the switching current as a function of electric field (*I*-*E*), from which we calculated the polarization-electric field (*P*-*E*) loops under various bending strains, where the primary strain direction is in-plane (parallel to surface in the <100>$_{pc}$ directions) and the electric field is out-of-plane (perpendicular to the surface) as shown in Figure 1b. Table 1 summarizes the applied radius of curvature (ROC; which is calculated by taking into account the bending radius and the distance of the neutral plane from the Kapton/ stage interface), the curvatures (1/ROC), and the corresponding calculated geometric in-plane (from curvature) and out-of-plane (via Poisson's ratio) strains. From the *I*-*E* (Fig. 2a) and *P*-*E* loops (Fig. 2b), key switching parameters, including the switching fields as well as the saturation and remnant polarizations, were extracted (Figs. 2c and 2d). These quantities are subsequently used to evaluate the energy storage density and efficiency at different uniaxial strain levels (Fig. 2e).

***Table 1***: *Radius of curvatures (ROC), curvatures, and calculated in-plane and out-of-plane strains.*

| ROC (mm) | 1/ROC ($mm^{-1}$) | Nominal In-plane strain | Nominal Out-of-plane strain |
|---|---|---|---|
| 0.566 | 1.767 | 9.54% | -3.53% |
| 0.866 | 1.154 | 6.23% | -2.31% |
| 5.066 | 0.197 | 1.07% | -0.39% |
| Flat | 0 | 0 | 0 |
| -3.534 | -0.282 | -1.53% | 0.57% |

As shown in Figs. 2a and 2b, under all applied strain conditions, the PZO membranes exhibit double *P*-*E* loops, along with four switching current peaks, which is characteristic of antiferroelectric behavior. As the in-plane uniaxial strain changes from compressive to tensile, the forward AFE-to-FE switching moves to higher electric field and the backward FE-to-AFE switching moves to lower fields, resulting in broader hysteresis loops. Meanwhile, the saturation polarization ($P_s$) gradually decreases with increasing in-plane tensile strain, whereas the remnant polarization at ($P_r$), remains close to 0 and nearly unchanged until the FE-to-AFE back-switching current peak shifts below 0V at the highest in-plane tensile strain (smallest ROC). To quantify the strain-induced evolution of switching characteristics, we define the fraction of back-switching polarization, i.e. $(P_s-P_r)/P_s$, where a value of 1 represents an ideal antiferroelectric behavior, and 0 represents an ideal ferroelectric behavior. For the highest in-plane tensile strain, the value of this ratio drops from 0.87 to 0.42, indicating a pronounced tendency towards a *dynamic* ferroelectric-like behavior, i.e. one where the metastable FE state does not have time to switch back to the AFE ground state within the measurement period [25]. As a consequence of this dependence of *P*-*E* loops on the strain conditions, both the energy storage density and the efficiency of the capacitors increase under compressive strains and decrease under tensile strains.

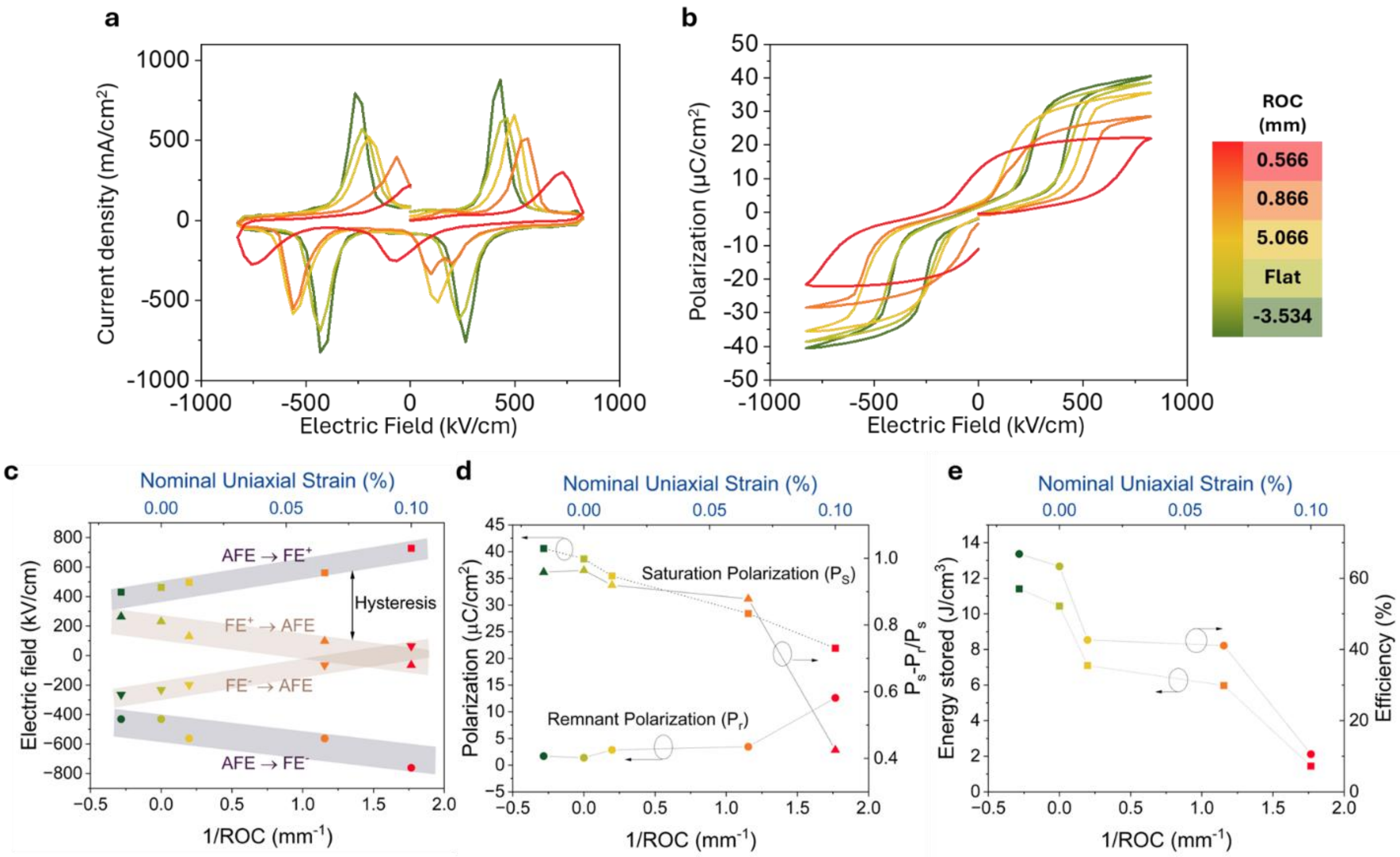

***Figure 2. Evolution AFE-FE switching hysteresis and functional properties of PZO flat membrane capacitors under uniaxial strains.*** *(a) Switching current-electric field (I-E) and (b) polarization-electric field (P-E) loops under different uniaxial strains. Evolution of (c) switching electric fields, (d) saturation polarization* $P_s$*, remnant polarization* $P_r$*, and the ratio* $(P_s - P_r)/P_s$ *versus curvature. (e) Energy storage density and efficiency as a function of curvature. Highlighted lines in (c) are guides for the eye. Top axes in (c-e) indicate the corresponding calculated nominal in-plane strains.*

## Atomistic simulations

The experimental results presented above demonstrate that strain effectively tunes the AFE-FE switching behavior of the PZO membranes, with key switching parameters exhibiting monotonic variations as the strain changes from compressive to the tensile. Previously, Guo et al. reported that the saturation polarization indeed decreases under in-plane tensile strain due to the electrostrictive coupling; however, they also reported that the switching hysteresis width *increases* under *compression*, which contrasts with the behavior observed in our study [21]. To elucidate the origin of this discrepancy and understand the microscopic mechanisms, we turn to atomistic simulations, utilizing a previously developed deep-learning interatomic potential [33]. These simulations reveal a bifurcated dependence of the hysteresis width on uniaxial strains, governed by the population of domain orientation variants. This analysis provides a comprehensive physical picture of the strain-tuning mechanisms in PZO membranes.

Figures 3a-c illustrate the three possible domain orientations of the antiferroelectric *Pbam* phase of PZO under combined uniaxial strain and electric field. In these configurations, the uniaxial strain (applied along the *y*-axis) and the electric field (applied along the *z*-axis) are orthogonal to each other, and both align with the pseudocubic axes, consistent with the experimental setup. The three possible orientations shown in Fig. 3a-c correspond to: (1) uniaxial strain along $[001]_{pc}$ with electric field along $[100]_{pc}$, (2) uniaxial strain along $[010]_{pc}$ with electric field along $[100]_{pc}$, and (3) uniaxial strain along $[100]_{pc}$ with electric field along $[001]_{pc}$. Orient-1 and Orient-2 both correspond to the $(120)_o$ domain, while Orient-3 corresponds to the $(001)_o$ domain. We note that Orient-2 and Orient-3 are energetically degenerate under uniaxial strain in the absence of electric field, whereas Orient-1 and Orient-2 become degenerate when an electric field is applied without strain constraints.

<u>*Individual effects of uniaxial strain and electric field*</u>

We first examine the orientation preference of the *Pbam* phase under different uniaxial strains at zero electric field. Our energy landscape analysis (Fig. 3d) reveals that the uniaxial compressive strain energetically favors Orient-1, while the uniaxial tensile strain favors Orient-2 and Orient-3. This strain-induced orientation preference arises from the anisotropy of the pseudocubic lattice parameters of the *Pbam* phase: the shorter pseudocubic axis preferentially aligns with the compressive strain, while the longer pseudocubic axis aligns with the uniaxial tensile strain.

This insight, gained from the energy landscape at 0 K, is further corroborated by stress evolution simulations at finite temperature. As shown in Fig. 3e, gradually changing the uniaxial stress from compressive to tensile induces a transition from Orient-1 to Orient-2 (or the equivalent Orient-3). Conversely, increasing compressive stress drives the transition from Orient-2 (or Orient-3) back to Orient-1. These results agree well with our XRD observations, which showed an increased relative population of the $(001)_o$ domain (corresponding to Orient-3) when the membrane capacitor is subjected to tensile bending compared with its flat state (Fig. 1c). Nevertheless, it should be noted that due to the limited size of the simulation supercell, the system remains single-domain throughout the simulation, and the reorientation proceeds through a homogeneous transformation. In practice, a homogeneous reorientation of an entire local region is energetically costly and a more plausible mechanism is domain-wall motion, involving the growth of favorably oriented domains at the expense of unfavorably oriented ones.

We then examine the effect of electric field by investigating the AFE-to-FE transition across different domain orientations in the absence of strain constraints. As shown in Fig. 3f, starting from the AFE state either in Orient-1 (equivalent to Orient-2 in the absence of strain constraint), or in Orient-3, gradually increasing the electric field induces a transition to the FE state at a specific electric field strength. Regardless of the initial AFE orientation, the induced FE phase is always the *R*3*c* phase (rhombohedral), with its symmetry lowered to *Cc* (monoclinic) because of the application of electric field. Notably, our simulation results indicate that Orient-3 requires a substantially higher electric field to induce the AFE-to-FE transition compared to Orient-1 and Orient-2, thus demonstrating that the AFE-to-FE transition field is strongly dependent on crystallographic orientation.

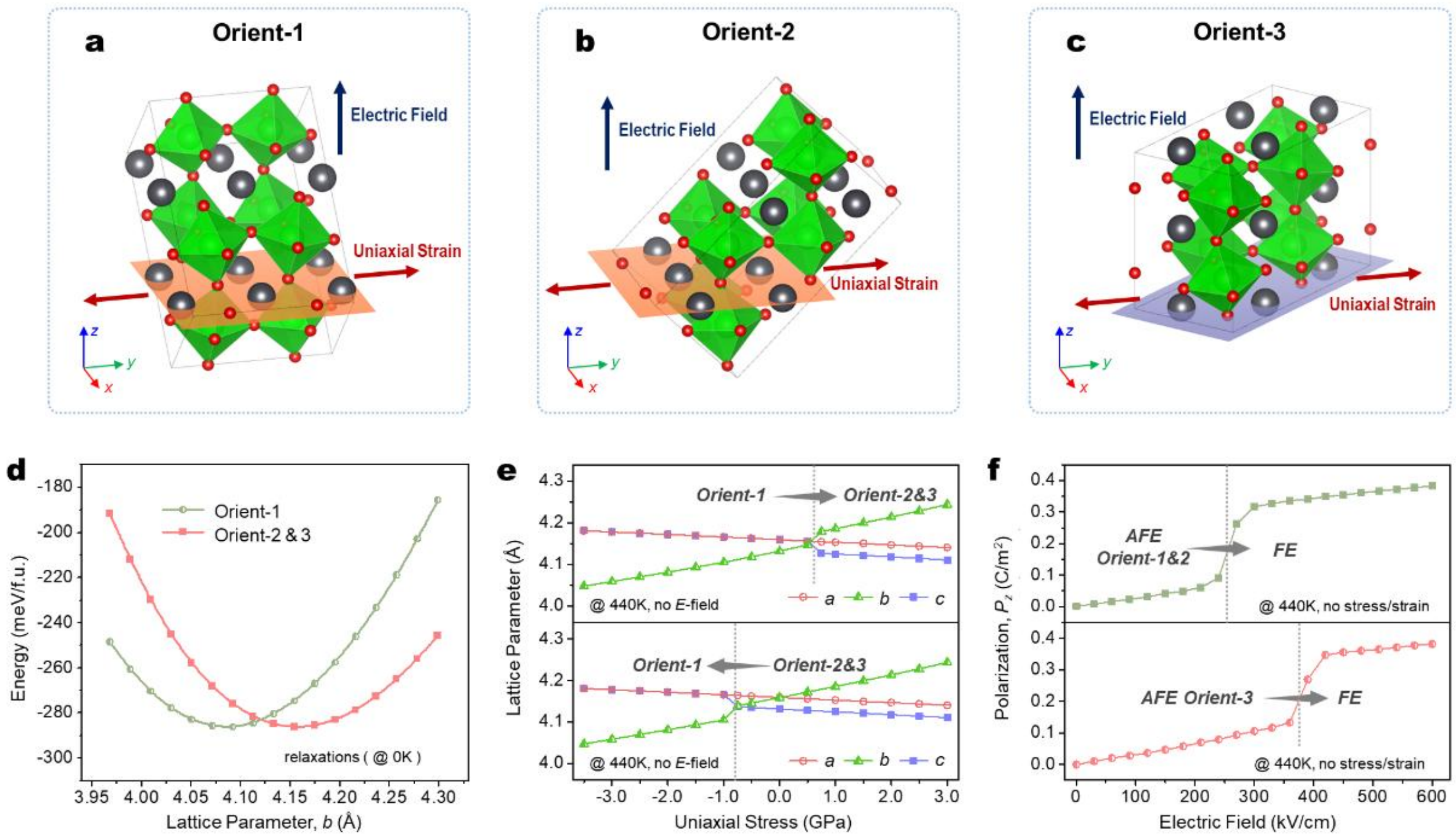

***Figure 3. Possible orientations of the AFE Pbam phase and the individual effects of uniaxial strain and electric field.*** *(a-c) Schematic illustrations of the three symmetry-inequivalent orientations of the AFE Pbam phase. The coordinate system is defined such that uniaxial strain is applied along the y-axis and the electric field is applied orthogonally along the z-axis. (d) Total energy of the relaxed Pbam phase under different uniaxial strains, where the strain is controlled by fixing the lattice parameter b along y-axis. (e) Evolution of lattice parameters under uniaxial stress at finite temperature (440 K), where a, b, and c correspond to pseudocubic lattice parameters along the x-, y-, and z-axes, respectively. Abrupt changes in lattice parameters indicate stress-induced orientation switching within the Pbam phase. Note that Orient-2 and Orient-3 are symmetry equivalent at zero electric field. (f) Electric-field-induced polarization at finite temperature (440 K) under zero external stress. Discontinuous jumps in polarization indicate the AFE-to-FE transitions. Note that under zero stress, Orient-1 and Orient-2 are symmetry*

*equivalent. Orient-3 requires a higher electric field to induce the AFE-to-FE transition compared with Orient-1 and Orient-2.*

*Combined effect of uniaxial strain and electric field*

Having clarified the individual roles of uniaxial strain and electric field, we proceed with investigating their combined effect by simulating the *P*-*E* loops under various uniaxial strain conditions (Fig. 4a and Fig. S3). Our simulations produce characteristic double *P*-*E* loops arising from reversible AFE-FE transitions across the examined strain range (Fig. 4b). In some of the *P*-*E* loops [e.g., Fig. 4(a1)], polarization plateaus are observed, which originate from the emergence of intermediate ferrielectric states characterized by unequal numbers of upward and downward dipoles (see Fig. S4). Similar phenomena have been reported experimentally, albeit without any external strains [34-38].

Notably, the hysteresis width of the simulated double *P*-*E* loops exhibits a pronounced dependence on the applied uniaxial strain. Our simulations reveal a "U-shaped" dependence: while the hysteresis width generally increases as the strain changes from compressive to tensile [Fig. 4(a3-a6)], which agrees with our experimental findings, a (smaller) widening of the hysteresis is also observed under large compressive strains [Fig. 4(a1-a2)]. This non-monotonic behavior appears to reconcile our results with those reported by Guo et al. [21], indicating that the trends depend both on the specific strain regime and domain population.

The strain dependence of the hysteresis width stems from the different evolution behavior for the transition fields of different domain variants. Figure 4(c1) summarizes the electric fields for the AFE-to-FE and FE-to-AFE transitions at different uniaxial strains, extracted from the simulated *P*-*E* loops. Under compressive strain, where the system prefers Orient-1, the electric field required to drive the transition from Orient-1 to the FE state increases with increasing compressive strain. Under tensile strain, where the system prefers Orient-2 and Orient-3, the transition fields for these variants also increase at larger tensile strain. Meanwhile, the field at which the FE state becomes unstable and reverts to AFE decreases continuously as strain changes from compressive to tensile, albeit at a slower rate than the increase of the AFE-FE field. The widening of the hysteresis at both large compressive and tensile strains is thus dominated by the increase of the forward AFE-to-FE switching field with different dominant domain variants [Fig. 4(c2)], but in the compressive strain case this widening is moderated by the (slower) increase of the FE-AFE field.

In addition to the intrinsic strain dependence of the transition field of individual AFE domain variants, the strain-induced changes in their relative populations also affect the switching hysteresis. Specifically, the increased population of Orient-3 under tensile strain, as evidenced experimentally in Fig. 1c, also contributes to increase in AFE-to-FE transition field, thereby further enlarging the hysteresis width. Consequently, the modulation of the switching behavior by uniaxial strain is attributed to the superposition of two mechanisms: (i) the intrinsic strain dependence of specific domain variants, and (ii) the extrinsic contribution arising from strain-induced changes in relative populations among different domain variants.

In contrast to the AFE phase, the FE $R3c$ phase possesses only one symmetry-inequivalent orientation. The properties associated with this FE phase, specifically the saturation polarization [Fig. 4(c3)] and the backward FE-to-AFE transition field [Fig. 4(c1)], both decrease continuously as the strain changes from compressive to tensile, aligning well with our experimental observations. The reduction in the apparent saturation polarization can be attributed to lattice elongation and polarization enhancement in the perpendicular direction, a consequence of the orthogonal configuration between the applied uniaxial strain and electric field. Meanwhile, the decrease in the backward switching field implies that the FE phase becomes increasingly stable under tensile strain, which can be associated with the volume effect, since the FE phase possesses a larger volume compared to the AFE phase [13, 27].

Finally, we note that the simulations occasionally exhibit asymmetric double hysteresis loops, most prominently under tensile strain [Figs. 4(a4-a6)]. Analysis of the structural evolution during electric-field cycling indicates that this asymmetry originates from a reorientation of the antiferroelectric (AFE) state recovered from the ferroelectric (FE) phase after electric-field unloading, which differs from the orientation of the initial AFE state. Under compressive strain, the zero-field AFE state consistently returns to Orient-1, whereas under tensile strain a random reorientation between the energetically degenerate Orient-2 and Orient-3 is observed (see Figs. S5 and S6). This reorientation can lead to a slight asymmetry between the positive and negative branches of the simulated hysteresis loops. Such asymmetric hysteresis is rarely observed in experiments, likely because macroscopic samples contain multiple domain orientations, whose coexistence effectively averages out these local asymmetries, resulting in nearly symmetric hysteresis loops typically observed in experiments.

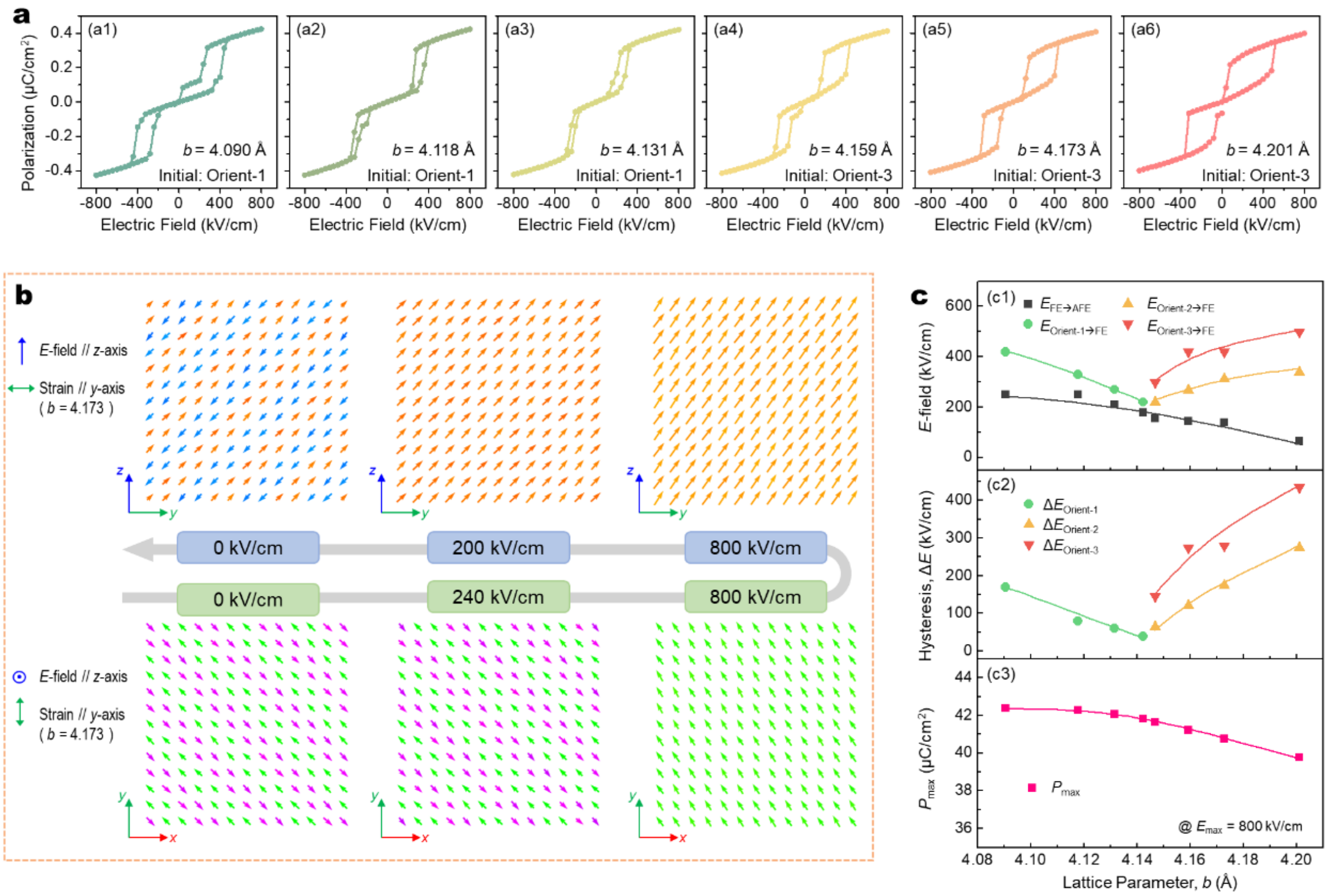

***Figure 4. Atomistic simulations of uniaxial strain effects on AFE-FE switching behavior.*** *(a) Double P-E loops under different uniaxial strain conditions at 440 K. The initial state prior to electric field application corresponds to the energetically favored orientation for each strain level (as annotated). The uniaxial strain was applied by fixing the lattice parameter b along y-axis. (b) Evolution of Pb displacements during the positive half-cycle of electric field for b = 4.173 Å. Arrows represent local Pb displacements relative to the average position of the 12 surrounding oxygen atoms, with colors indicating the displacement directions. The application of positive half-cycle of electric field changes the initial Orient-3 to Orient-2, which are energetically degenerate at zero electric field. (c) The AFE-FE switching electric fields, switching hysteresis, and maximum polarization as functions of the lattice parameter b. Lines in (c) are guides for the eye. All simulations were performed using 12×12×12 supercells.*

## Summary and Conclusion

Enabled by a robust gold-bonding transfer method, our experiments demonstrate that the substrate bending provides an effective way to dynamically modulate the strain and, with it, the AFE-FE switching behavior. Corroborated by atomistic simulations, our results reveal that the strain-dependent evolution of the hysteresis loop arises from the superposition of two mechanisms: a strain dependence of the switching behavior of individual domain variants, and a strain-induced redistribution of domain populations. Depending on the dominant domain variants, the hysteresis width can exhibit different strain-dependent trends. These insights reconcile apparently discrepant experimental observations, and establish a unified microscopic understanding needed for tailoring antiferroelectric membranes for flexible electronic applications.

## Acknowledgements

ICN2 is supported by the Severo Ochoa program from Spanish MCIN / AEI (Grant No. CEX2021-001214-S. U.S. acknowledges the funding for this project by Grant No. PID2019-108573GB-C21 (FOx-Me) funded by the Spanish Ministry of Science and Innovation. H.Z. acknowledges support from the National Natural Science Foundation of China (Grant No. 12674116), and the Fundamental Research Funds for the Central Universities of China (Grant No. 104972026KFYjc0115). D.P. acknowledges the "Ramon y Cajal" Senior Research Fellowship (RYC-2023-044643-I) from the Spanish Ministry of Science, Innovation and Universities. and grant PID2022-140589NB-I00, funded by MCIN/AEI/10.13039/501100011033. F.G.O. acknowledges financial support from the Fonds de la Recherche Scientifique (FNRS) through the grant FNRS-CR 1.B.227.25F. This work is supported by the European Union's Horizon 2020 research and innovation programme under Grant Agreement No. 964931 (TSAR), and by F.R.S.-

FNRS Belgium under PDR Grants No. T.0107.20 (PROMOSPAN) and No. T.0128.25 (TOPOTEX). The authors acknowledge the use of the CECI supercomputer facilities funded by the F.R.S.-FNRS (Grant No. 2.5020.1) and the Tier-1 supercomputer of the Fédération Wallonie-Bruxelles funded by the Walloon Region (Grant No. 1117545). GC acknowledges support from the EU through HORIZON-MSCA-SE-3D-TOPO action (Grant agreement ID: 101236483) and from Spain's MICINN through grant PID2023-148673NB-I00 (Project GRIPHO2).

**Associated content: Supplementary Information** includes:

S1. Method details

*S1.1. Experimental details*

*S1.2. Calculation details*

S2. Estimation of strain gradient induced flexoelectric field

S3. XRD peak fittings for the bent SRO/PZO/SRO capacitors

S4. Supplementary analysis of the simulated double *P*-*E* hysteresis loops

References

**Data availability:** All data collected during this study is available from the authors upon reasonable request.

**Reference**

[1] C. Kittel, Theory of antiferroelectric crystals, Physical Review, 82, 729-732 (1951).

[2] E. Sawaguchi, H. Maniwa, S. Hoshino, Antiferroelectric structure of lead zirconate, Physical Review 83, 1078-1078 (1951).

[3] G. Shirane, E. Sawaguchi, Y. Takagi, Dielectric properties of lead zirconate, Physical Review 84, 476-481 (1951).

[4] M.E. Lines, A.M. Glass, Principles and Applications of Ferroelectrics and Related Materials, Oxford University Press, 1977.

[5] K.M. Rabe, Antiferroelectricity in oxides: A reexamination, in Functional Metal Oxides, edited by S.B. Ogale, T.V. Venkatesan, and M.G. Blamire, John Wiley & Sons, 2013.

[6] C.A. Randall, Z. Fan, I. Reaney, L.-Q. Chen, S. Trolier-McKinstry, Antiferroelectrics: History, fundamentals, crystal chemistry, crystal structures, size effects, and applications, Journal of the American Ceramic Society, 104, 3775-3810, (2021).

[7] X. Hao, J. Zhai, L.B. Kong, Z. Xu, A comprehensive review on the progress of lead zirconate-based antiferroelectric materials, Progress in Materials Science, 63, 1-57 (2014).

[8] A.S. Mischenko, Q. Zhang, J.F. Scott, R.W. Whatmore, N.D. Mathur, Giant electrocaloric effect in thin-film $PbZr_{0.95}Ti_{0.05}O_3$, Science, 311, 1270-1271 (2006).

[9] B. Yang, Y. Liu, R.-J. Jiang, S. Lan, S.-Z. Liu, Z. Zhou, L. Dou, M. Zhang, H. Huang, L.-Q. Chen, Y.-L. Zhu, S. Zhang, X.-L. Ma, C.-W. Nan, Y.-H. Lin, Enhanced energy storage in antiferroelectrics via antipolar frustration, Nature, 637, 1104-1110 (2025).

[10] C. Liu, Y. Si, H. Zhang, C. Wu, S. Deng, Y. Dong, Y. Li, M. Zhuo, N. Fan, B. Xu, P. Lu, L. Zhang, X. Lin, X. Liu, J. Yang, Z. Luo, S. Das, L. Bellaiche, Y. Chen, Z. Chen, Low voltage driven high-performance thermal switching in antiferroelectric $PbZrO_3$ thin films, Science, 382, 1265-1269 (2023).

[11] Y.-X. Wu, Z. Tang, F. Qiu, Z.-J. Chen, J.-L. Fang, Y.-P. Jiang, X.-G. Tang, S.-F. Li, L. Ma, Y.-C. Zhou, J. Gao, The reverse diode behavior, negative photoconductivity effect and carrier transport mechanisms in $PbZrO_3$ antiferroelectric memristor for neuromorphic computing, Advanced Functional Materials, 36, e13366 (2026).

[12] Z. Lv, H. Wang, J. Cao, C. Zhang, G. Zhao, R. Yu, B. Zhang, X. Xu, Y. Jiang, J. Miao, Tunable volatile to non-volatile resistive switching in $PbZrO_3$ antiferroelectric thin film for neuromorphic computing, Advanced Materials Interfaces, 9, 2201005 (2022).

[13] X. Tan, C. Ma, J. Frederick, S. Beckman, K.G. Webber, The antiferroelectric ↔ ferroelectric phase transition in lead-containing and lead-free perovskite ceramics, Journal of the American Ceramic Society, 94, 4091-4107 (2011).

[14] Y. Si, T. Zhang, C. Liu, S. Das, B. Xu, R.G. Burkovsky, X.-K. Wei, Z. Chen, Antiferroelectric oxide thin-films: Fundamentals, properties, and applications, Progress in Materials Science, 142, 101231 (2024).

[15] G. Catalan, A. Gruverman, J. Íñiguez-González, D. Meier, M. Trassin, A modern perspective on antiferroelectrics, Nature Materials, 25, 557-565 (2026).

[16] Pushan Ayyub, Soma Chattopadhyay, R. Pinto, and M. S. Multani, Ferroelectric behavior in thin films of antiferroelectric materials, Physical Review B, 57, R5559-R5562 (1998).

[17] Ayan Roy Chaudhuri, Miryam Arredondo, Angelika Hähnel, Alessio Morelli, Michael Becker, Marin Alexe, and Ionela Vrejoiu, Epitaxial strain stabilization of a ferroelectric phase in $PbZrO_3$ thin films, Physical Review B, 84, 054112 (2011).

[18] D. Lu, D.J. Baek, S.S. Hong, L.F. Kourkoutis, Y. Hikita, H.Y. Hwang, Synthesis of freestanding single-crystal perovskite films and heterostructures by etching of sacrificial water-soluble layers, Nature Materials, 15, 1255-1260 (2016).

[19] D. Pesquera, A. Fernández, E. Khestanova, L.W. Martin, Freestanding complex-oxide membranes, Journal of Physics: Condensed Matter, 34, 383001 (2022).

[20] Y. Li, E. Zatterin, M. Conroy, A. Pylypets, F. Borodavka, A. Björling, D.J. Groenendijk, E. Lesne, A.J. Clancy, M. Hadjimichael, D. Kepaptsoglou, Q.M. Ramasse, A.D. Caviglia, J. Hlinka, U. Bangert, S.J. Leake, P. Zubko, Electrostatically driven polarization flop and strain-induced curvature in free-standing ferroelectric superlattices, Advanced Materials, 34, 2106826 (2022).

[21] Y. Guo, B. Peng, G. Lu, G. Dong, G. Yang, B. Chen, R. Qiu, H. Liu, B. Zhang, Y. Yao, Y. Zhao, S. Li, X. Ding, J. Sun, M. Liu, Remarkable flexibility in freestanding single-crystalline antiferroelectric $PbZrO_3$ membranes, Nature Communications, 15, 4414 (2024).

[22] R. Xu, K.J. Crust, V. Harbola, R. Arras, K.Y. Patel, S. Prosandeev, H. Cao, Y.-T. Shao, P. Behera, L. Caretta, W.J. Kim, A. Khandelwal, M. Acharya, M.M. Wang, Y. Liu, E.S. Barnard, A. Raja, L.W. Martin, X.W. Gu, H. Zhou, R. Ramesh, D.A. Muller, L. Bellaiche, H.Y. Hwang, Size-induced ferroelectricity in antiferroelectric oxide membranes, Advanced Materials, 35, 2210562 (2023).

[23] R. Xu, J. Huang, E.S. Barnard, S.S. Hong, P. Singh, E.K. Wong, T. Jansen, V. Harbola, J. Xiao, B.Y. Wang, S. Crossley, D. Lu, S. Liu, H.Y. Hwang, Strain-induced room-temperature ferroelectricity in $SrTiO_3$ membranes, Nature Communications, 11, 3141 (2020).

[24] G. Dong, S. Li, M. Yao, Z. Zhou, Y.-Q. Zhang, X. Han, Z. Luo, J. Yao, B. Peng, Z. Hu, H. Huang, T. Jia, J. Li, W. Ren, Z.-G. Ye, X. Ding, J. Sun, C.-W. Nan, L.-Q. Chen, J. Li, M. Liu, Super-elastic ferroelectric single-crystal membrane with continuous electric dipole rotation, Science, 366, 475-479 (2019).

[25] U. Saeed, D. Pesquera, Y. Liu, I. Fina, S. Ganguly, J. Santiso, J. Padilla-Pantoja, J.M. Caicedo Roque, X. Liao, G. Catalan, Switching dynamics and improved efficiency of free-standing antiferroelectric capacitors, Advanced Electronic Materials, 10, 2400102 (2024).

[26] J. Wang, S. Xu, S. Meyer, S. Wu, S. Bandyopadhyay, X. He, Q. Miao, S. Huang, P. Li, K. Zhao, E.-J. Guo, C. Ge, B. Dupé, P. Ghosez, K. Chang, K. Jin, Manipulation of ferroic orders via continuous biaxial strain engineering in multiferroic bismuth ferrite, Advanced Science, 12, 2417165 (2025).

[27] H. Pan, M. Zhu, E. Banyas, L. Alaerts, M. Acharya, H. Zhang, J. Kim, X. Chen, X. Huang, M. Xu, I. Harris, Z. Tian, F. Ricci, B. Hanrahan, J.E. Spanier, G. Hautier, J.M. LeBeau, J.B. Neaton, L.W. Martin, Clamping enables enhanced electromechanical responses in antiferroelectric thin films, Nature Materials, 23, 944-950 (2024).

[28] J. Frederick, X. Tan, W. Jo, Strains and polarization during antiferroelectric-ferroelectric phase switching in $Pb_{0.99}Nb_{0.02}[(Zr_{0.57}Sn_{0.43})_{1-y}Ti_y]_{0.98}O_3$ ceramics, Journal of the American Ceramic Society, 94, 1149-1155 (2011).

[29] M.M.V. Taklo, P. Storås, K. Schjølberg-Henriksen, H.K. Hasting, H. Jakobsen, Strong, high-yield and low-temperature thermocompression silicon wafer-level bonding with gold, Journal of Micromechanics and Microengineering, 14, 884 (2004).

[30] G.G. Zhang, X.F. Ang, Z. Chen, C.C. Wong, J. Wei, Critical temperatures in thermocompression gold stud bonding, Journal of Applied Physics, 102, 063519 (2007).

[31] P. Vales-Castro, K. Roleder, L. Zhao, J.-F. Li, D. Kajewski, G. Catalan, Flexoelectricity in antiferroelectrics. Applied Physics Letters 113, 132903 (2018).

[32] P. Dufour, Structural and physical properties of antiferroic oxide thin films, Ph.D. Dissertation, Université Paris-Saclay, 2023.

[33] H. Zhang, H.-C. Thong, L. Bastogne, C. Gui, X. He, P. Ghosez, Finite-temperature properties of the antiferroelectric perovskite $PbZrO_3$ from a deep-learning interatomic potential, Physical Review B, 110, 054109 (2024).

[34] Y.-P. Feng, M.-X. Zhu, R.-J. Jiang, Y.-J. Wang, Y.-L. Tang, Y.-L. Zhu, X.-L. Ma, Direct observation of electric-field-driven phase transitions associated with energy storage in antiferroelectric films, Advanced Science, 13, e17897 (2026).

[35] M. Zhu, M. Xu, L. Alaerts, H. Pan, C. Gilgenbach, G. Hautier, L.W. Martin, J.M. LeBeau, Direct determination of antiferroelectric-to-ferroelectric phase transition pathways in $PbZrO_3$ with Operando Electron Microscopy, arXiv: 2509.07194.

[36] M. Zhu, M. Xu, H. Pan, C. Gilgenbach, L.W. Martin, J.M. LeBeau, In situ atomic-scale imaging unveils clamping-enhanced strain and phase transition pathways in antiferroelectric thin films, Microscopy and Microanalysis, 31, ozaf048.560 (2025).

[37] Z. Fu, X. Chen, Z. Li, T. Hu, L. Zhang, P. Lu, S. Zhang, G. Wang, X. Dong, F. Xu, Unveiling the ferrielectric nature of $PbZrO_3$-based antiferroelectric materials, Nature Communications, 11, 3809 (2020).

[38] Z. Li, Z. Fu, H. Cai, T. Hu, Z. Yu, Y. Luo, L. Zhang, H. Yao, X. Chen, S. Zhang, G. Wang, X. Dong, F. Xu, Discovery of electric devil's staircase in perovskite antiferroelectric, Science Advances, 8, eabl9088 (2022).

Supplementary Materials for

# **Uniaxial strain tuning of functional properties in antiferroelectric $PbZrO_3$ membrane capacitors**

*Umair Saeed[1,2,3*‡], Huazhang Zhang[4,5*‡], David Pesquera[1], Fernando Gómez-Ortiz[5], Jessica Padilla- Pantoja[1], Jose Santiso[1], Philippe Ghosez[5], Gustau Catalan[1,6]*

Corresponding authors: ‡ umasa@dtu.dk

‡ zhanghz782@126.com

**This file includes:**

## S1. Method details

### *S1.1. Experimental details*

The epitaxial stack was grown using Pulsed Laser Deposition (PLD) using a KrF excimer laser (248 nm, COMPex 102, Lambda Physik) on $SrTiO_3$ (001) substrate from CrysTec. First, 5 nm $Sr_3Al_2O_6$ layer was grown using laser fluence of 2.0 J cm$^{-2}$ at 750°C in dynamic oxygen partial pressure of 1 mTorr with 1 Hz laser frequency. The bottom $SrRuO_3$ layer was grown with a laser fluence of 1.25 J cm$^{-2}$ at 700°C in dynamic oxygen partial pressure of 100 mTorr at 2 Hz frequency. $PbZrO_3$ and top $SrRuO_3$ layers were then grown with laser fluence of 1.5 J cm$^{-2}$ at 450°C with dynamic oxygen partial pressure of 100 mTorr, with laser frequency of 2 Hz for $PbZrO_3$ and 1 Hz for $SrRuO_3$. The stack was then cooled down in 100 Torr of static oxygen pressure at 5°C min$^{-1}$.

After gold bonding the membrane to the Kapton substrate, the top SRO electrodes were patterned via photolithography. The top SRO layer was spin-coated with the photoresist AZ5214E from Laurell Tech, using WS400BZ-6NPP LITE. The program for spin coating was 500 rotations per minute (rpm) for 5 s and then 4000 rpm for 55 s. The pattern was designed using the software Clewin and the patterning was done using DWL Lithography using Kloé Dilase 250. The electrodes of diameter 50 µm were designed with the patterning velocity of the UV laser 0.1 µm s$^{-1}$ with a laser modulation of 10. After the first UV exposure, the sample was baked at 120°C for 120 s and then again exposed completely to UV using the Kloé KLUB 3 for 10 s. The photoresist was then developed by AZ 726 and the SRO was etched using sodium periodate ($NaIO_4$) chemical etchant. Finally, the sample was washed with acetone to remove all remaining photoresist.

The XRD was measured using Panalytical X'pert Pro diffractometer (Copper K$\alpha_1$, 1.540598 Å), using a hybrid 2-bounce primary monochromator on the incident beam side and a PIXcel line detector.

The functional properties of the capacitors were measured using Radiant Precision LC wherein the displacement current was measured as a function of voltage cycling at a frequency of 2 kHz, and I-E and P-E loops were calculated.

*S1.2. Calculation details*

The atomistic simulations were performed based on a previous developed deep-learning interatomic potential [S1]. The model was trained using DeePMD-kit [S2] and based on a first-principles dataset containing approximately $1.2\times10^4$ configurations calculated using ABINIT software [S3] and GGA-PBEsol functional [S4]. The model has been extensively validated and demonstrated to well capture the complex potential energy landscape of PZO, and correctly reproduce the temperature dependent phase transitions and the antiferroelectric double *P*-*E* hysteresis loops.

Structural relaxations under uniaxial strain and electric fields were performed using the Broyden-Fletcher-Goldfarb-Shanno minimization algorithm implemented in ASE [S5]. The uniaxial strain was applied by fixing the lattice parameter along the strain axis. The electric field was applied by adding electric forces on each atom according to its cubic-phase Born effective charge tensors, which were obtained from the density functional perturbation theory calculations using ABINIT.

Finite-temperature molecular dynamics simulations were carried out using LAMMPS [S6] on 12×12×12 supercells using NPT ensemble. The temperature and pressure were controlled by a Nose-Hoover thermostat and barostat, with an integration time step for the MD simulations of 1 fs. Since the interatomic potential over-stabilizes the ferroelectric $R3c$ phase by about 5.0 meV/f.u. [S1], a simulation temperature of 440 K (higher than the room temperature) was selected, so as to reproduce the antiferroelectric double *P*-*E* hysteresis loops. To simulate the *P*-*E* loops, a bipolar triangular electric field waveform with the amplitude of $E_{max}$ = 800 kV/cm was applied. At each electric field step, the simulations were run for on less than 80 ps. The first half of the trajectory was used for thermalization and equilibration, while the second half was used to calculate the average structure and analyze properties.

## S2. Strain calculations and estimation of strain gradient induced flexoelectric field

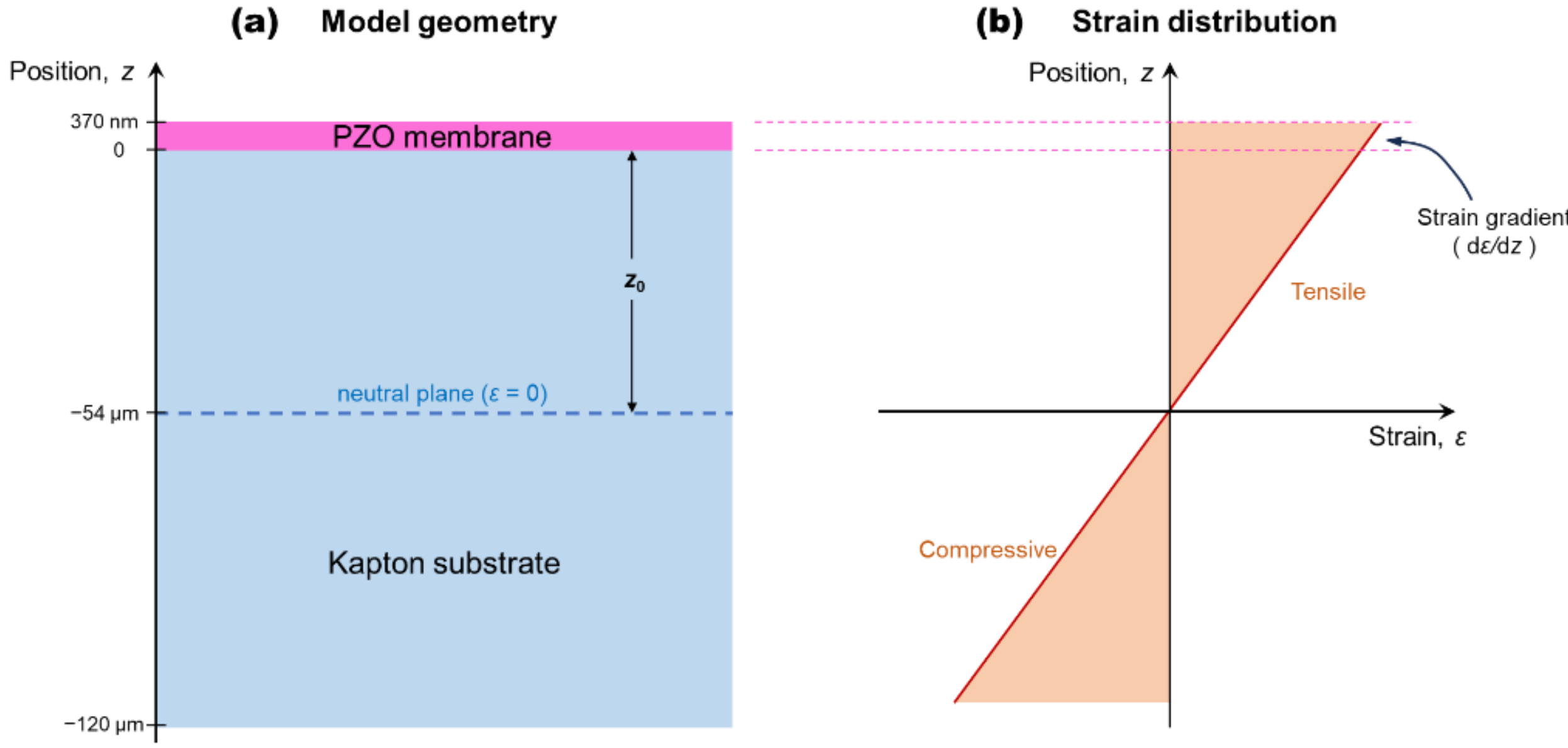

***Figure S1.*** *Schematic illustrations of the (a) geometry and (b) strain distribution across the PZO/Kapton bilayer model.*

For simplicity, we consider a bilayer model (Fig. S1a) consisting of a PZO membrane on a Kapton substrate. Assuming perfect bonding at the interface with no relative slip during bending, the position of the neutral plane can be calculated as [S7]

$$z_0 = \frac{Y_s t_s^2 - Y_f t_f^2}{2(Y_s t_s + Y_f t_f)}$$

where $t_s$ and $t_f$ are the thicknesses, and $Y_s$ and $Y_f$ are the Young's moduli of the substrate and film, respectively.

In our case, the thickness of the Kapton tape is 120 μm compared to the 370 nm capacitor thickness. The Young's modulus for PZO is 217 GPa and for Kapton is 2.76 GPa [S8]. The neutral plane, hence, lies about 54 μm away from the membrane-Kapton interface, inside the Kapton substrate.

The distance of the Kapton-stage interface from the neutral plane then comes out to be 120 – 54 = 66 μm which can then be added to the radius of the protrusion of the stage to obtain ROC. The nominal in-plane strain, ε for each case can be calculated by:

$$\varepsilon = \frac{z_0}{ROC}$$

Subsequently, nominal out-of-plane strain is then calculated using Poisson's ratio ($v$):

$$\varepsilon_{OOP} = \varepsilon * v$$

At the smallest experimental bending curvature radius $r_{\min} = 0.5 + (0.120 - 0.054) = 0.566$ mm, the effective strain gradient is

$$\frac{\mathrm{d}\varepsilon}{\mathrm{d}z} = \frac{1}{r_{\min}} = 1766.8\ \mathrm{m}^{-1}$$

Then, the resulting flexoelectric field $E_{\mathrm{flexo}}$ and flexoelectric polarization $P_{\mathrm{flexo}}$ can be estimated as

$$E_{\mathrm{flexo}} = f \cdot \frac{\mathrm{d}\varepsilon}{\mathrm{d}z} = 0.035\ \mathrm{kV/cm}$$

$$P_{\mathrm{flexo}} = \mu \cdot \frac{\mathrm{d}\varepsilon}{\mathrm{d}z} = 6.9 \times 10^{-4}\ \mu\mathrm{C/cm}^2$$

in which the typical values of flexocoupling coefficient ($f \approx 2.0$ V) and flexoelectric coefficient ($\mu \approx 3.9$ nC/m) are used [S9].

As seen, the estimated flexoelectric field is several orders of magnitude smaller than the typical value of the applied electric field and the switching electric fields observed in this work. Similarly, the flexoelectricity induced polarization is negligible compared to the measured polarization of PZO. Therefore, the flexoelectric effect contributes negligibly to the AFE-FE switching in our PZO membranes.

**S3. XRD peak fittings for the bent SRO/PZO/SRO capacitors**

To calculate the relative populations of the two domain variants in the bent PZO layer with a radius of 5.0 mm, and since, the peaks may be convoluted with the additional peaks from the copper stage and gold, the $\theta$-$2\theta$ scan was fitted with multiple pseudo-Voigt peaks, as shown in Fig. S2. The relative domain populations can be calculated using the relative intensities of the two fitted peaks. The resulting population ratio of $(240)_o$ to $(004)_o$ equals 1 : 0.6.

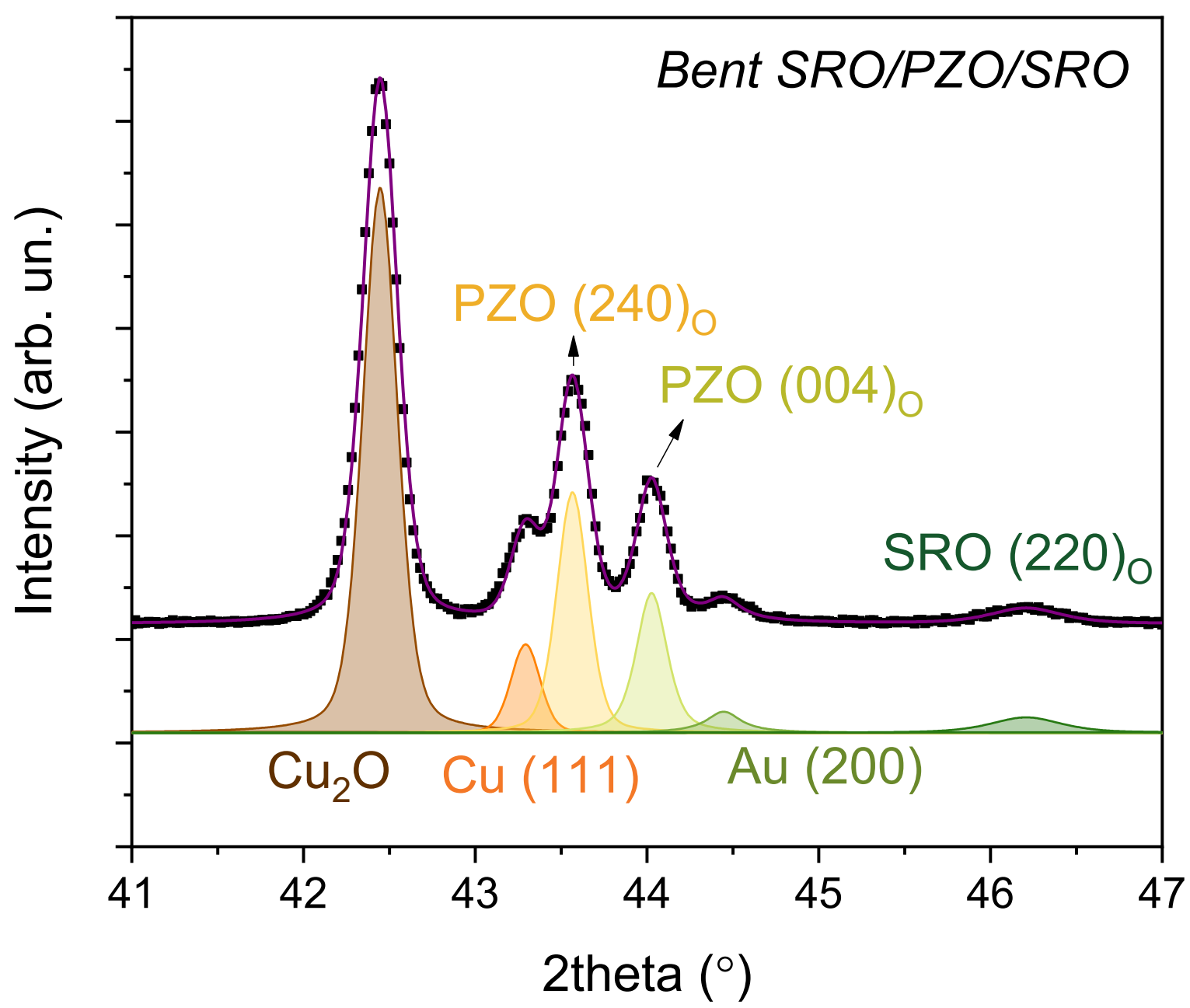

***Figure S2.*** *XRD peak fittings for the* $\theta$-$2\theta$ *scan of the SRO/PZO/SRO membrane bent with a radius of 5.0 mm, the dots are the data points, and the solid line is the total fit. The fitted peaks are vertically offset for better visualization.*

## S4. Supplementary analysis of the simulated double *P-E* hysteresis loops

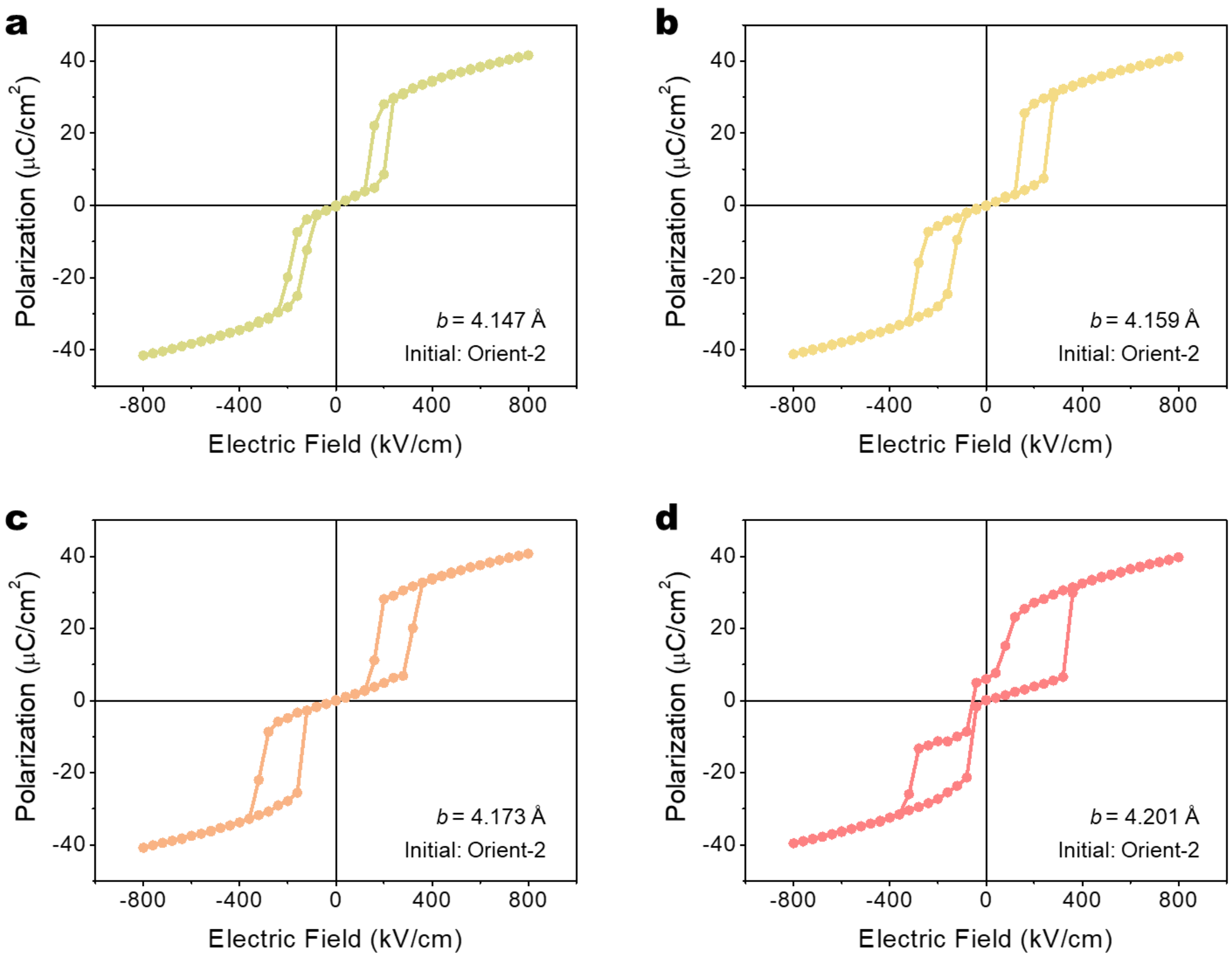

***Figure S3.*** *Simulated double P-E hysteresis loops with Orient-2 as the initial state under various uniaxial strains at 440 K. The pseudocubic lattice parameter along the uniaxial strain direction is fixed to (a) b = 4.147 Å, (b) b = 4.159 Å, (c) b = 4.173 Å and (d) b = 4.201 Å.*

Figure S3 presents the simulated double *P-E* hysteresis loops initialized from Orient-2 as the initial state under various uniaxial strains. Corresponding results for Orient-1 and Orient-3 are shown in Fig. 4a of the main text. For Orient-2, it is observed that the hysteresis width gradually broadens with increasing tensile strain. This trend is similar for the Orient-3, whereas an opposite trend is observed for the Orient-1.

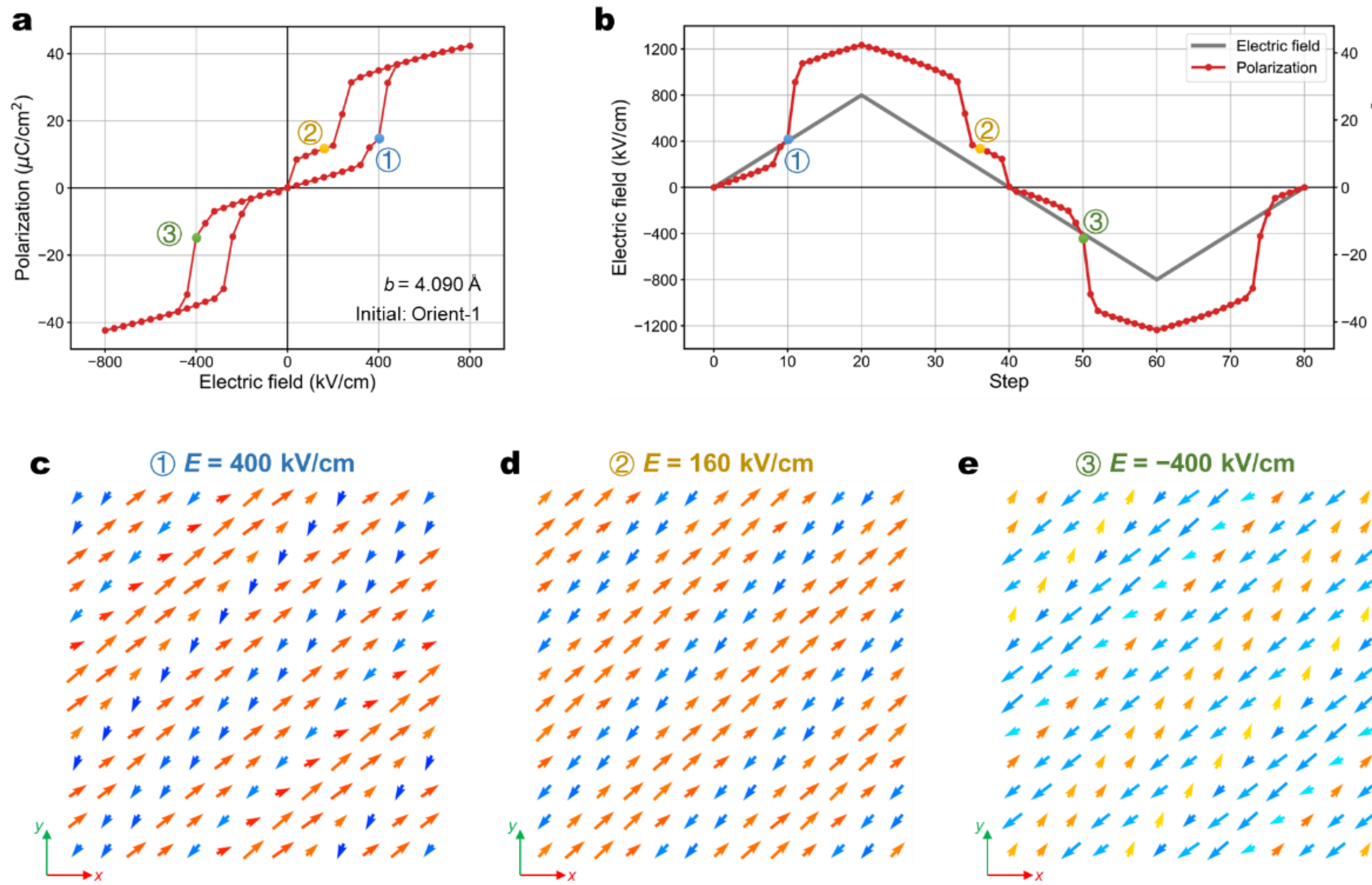

***Figure S4.*** *Ferrielectric intermediate structures observed during the electric field cycling. (a) Double P-E hysteresis loop. (b) Waveforms of polarization and electric field. (c-d) Representative Pb displacement maps of the ferrielectric intermediate structures. The results correspond to Fig. 4(a1) of the main text.*

Under certain uniaxial strain conditions, polarization plateaus appear on the double *P*-*E* loops, as a typical example shown in Fig. S4. By visualizing the corresponding Pb displacement maps, it turns out that the polarization plateaus arise from the emergence of intermediate ferrielectric structures characterized by unequal numbers of upward and downward dipoles.

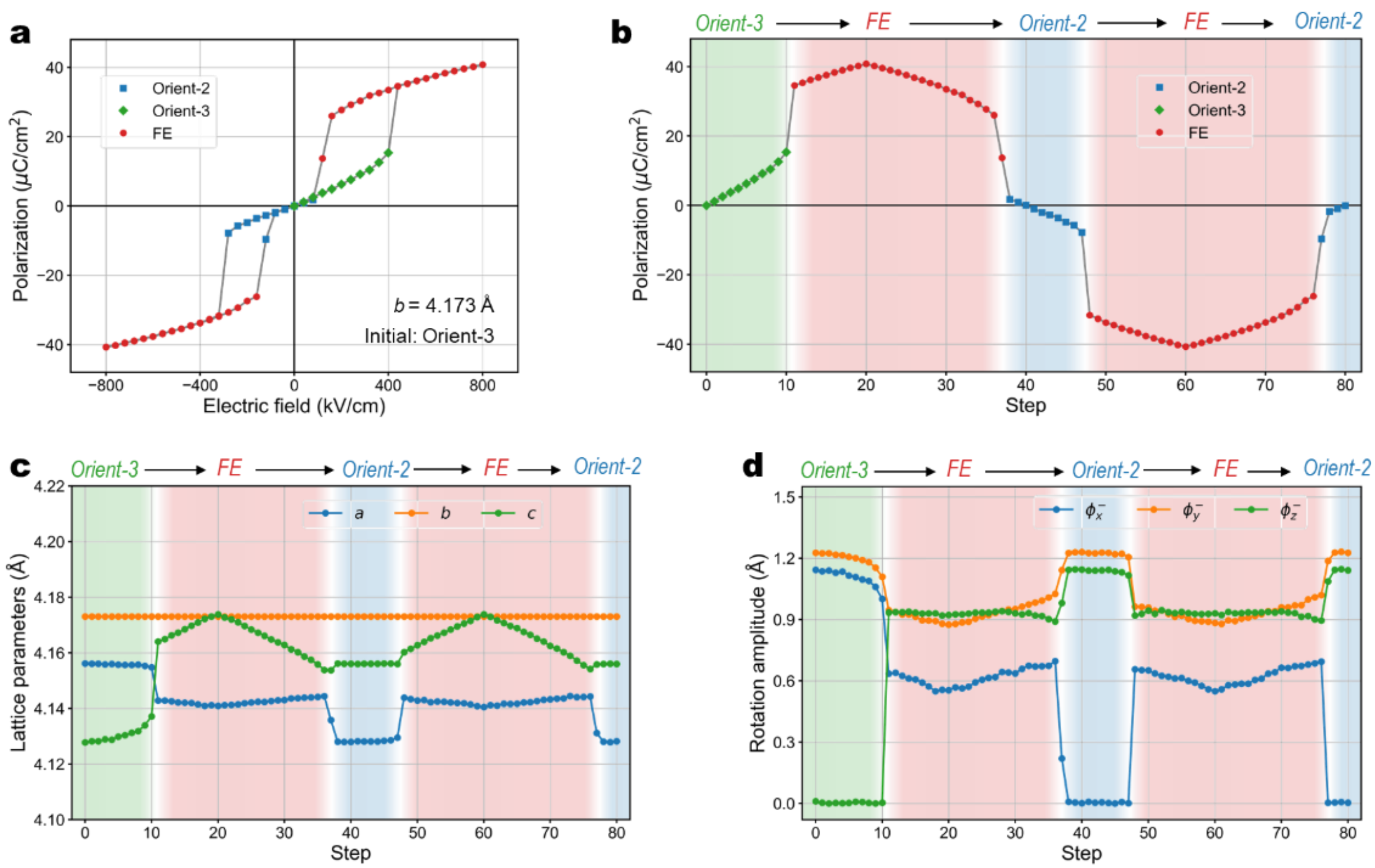

***Figure S5.*** *Asymmetric P-E hysteresis loop arising from electric field induced reorientation of antiferroelectric state. (a) Double P-E hysteresis loop. (b) Polarization curve. Different symbols in (a) and (b) represent different states, as indicated in the legend. (c) Evolution of pseudocubic lattice parameters a, b, and c along the x-, y-, and z-axes, respectively. (c) Evolution of antiphase octahedral rotations* $\phi_x^-$, $\phi_y^-$, *and* $\phi_z^-$ *about the x-, y-, and z-axes, respectively. The results correspond to Fig. 4(a5) of the main text.*

Figure S5 displays a representative asymmetric double *P-E* hysteresis loop, simulated starting from the Orient-3 state under tensile strain. Notably, the hysteresis width of the positive branch is larger than that of the negative branch. As shown in Figs. S5(c) and (d), the initial AFE state is in Orient-3, whereas the AFE state recovered from FE state upon removal of the electric field is in Orient-2. According to Fig. 3(f) of the main text, the electric field required to induce the transition from AFE Orient-3 to FE is substantially higher than that for Orient-2. Therefore, the observed asymmetry in the double *P-E* hysteresis loop arises directly from this field-induced reorientation of the AFE state.

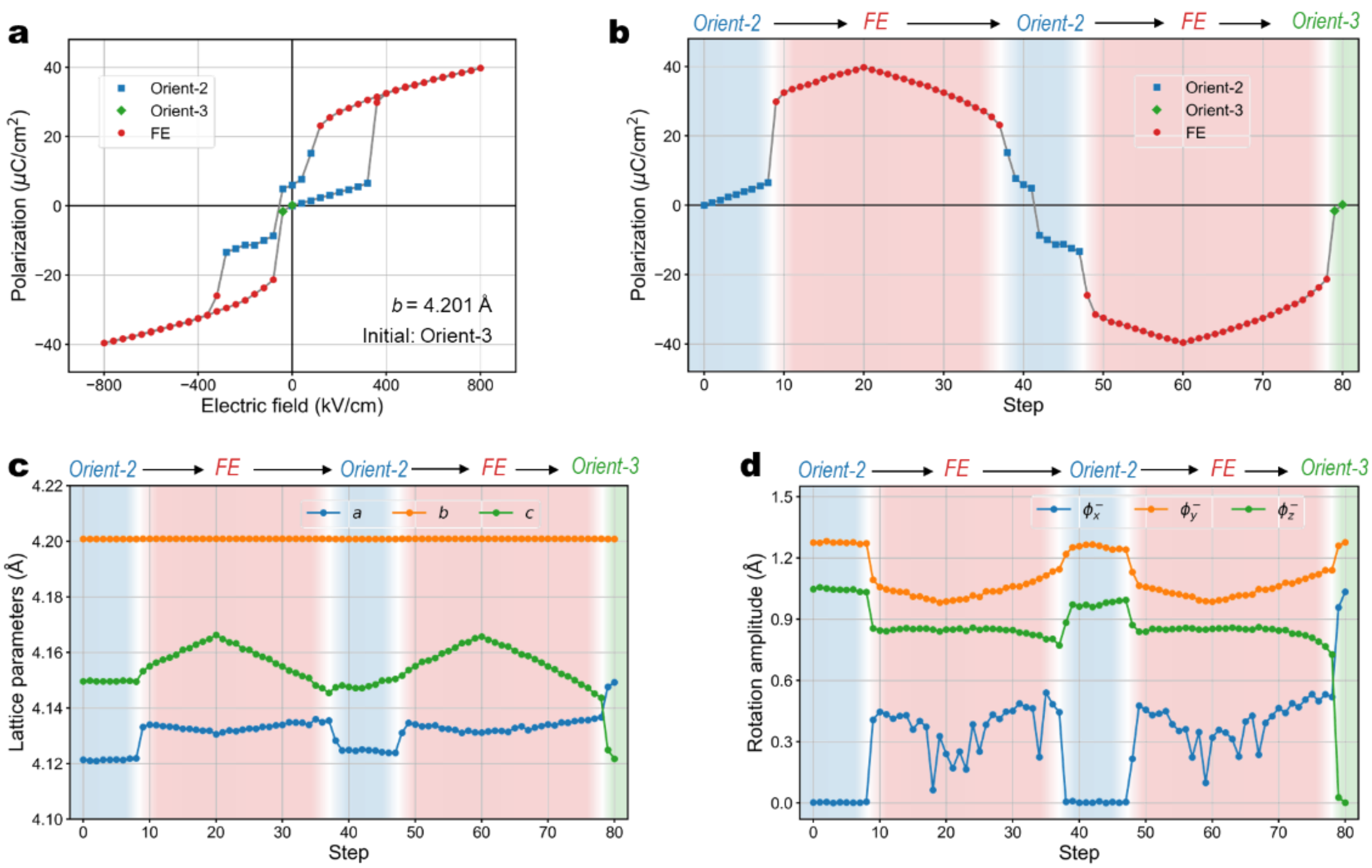

***Figure S6.*** *Example of a P-E hysteresis loop initialized with Orient-2 and finalized with Orient-3. (a) P-E hysteresis loop. (b) Polarization curve. Different symbols in (a) and (b) represent different states, as indicated in the legend. (c) Evolution of pseudocubic lattice parameters a, b, and c along the x-, y-, and z-axes, respectively. (c) Evolution of antiphase octahedral rotations* $\phi_x^-$, $\phi_y^-$, *and* $\phi_z^-$ *about the x-, y-, and z-axes, respectively. The results correspond to Fig. S3(d).*

Although the orientation changes from Orient-3 to Orient-2 after electric-field exposure are more frequently observed in our simulations, the reverse process, i.e., the orientation change from Orient-2 to Orient-3, can also occur. An example of this reverse transition is provided in Fig. S6.

**Reference**

[S1] H. Zhang, H.-C. Thong, L. Bastogne, C. Gui, X. He, P. Ghosez, Finite-temperature properties of the antiferroelectric perovskite $PbZrO_3$ from a deep-learning interatomic potential, Physical Review B 110 (2024) 054109.

[S2] J. Zeng, D. Zhang, D. Lu, P. Mo, Z. Li, Y. Chen, M. Rynik, L.a. Huang, Z. Li, S. Shi, Y. Wang, H. Ye, P. Tuo, J. Yang, Y. Ding, Y. Li, D. Tisi, Q. Zeng, H. Bao, Y. Xia, J. Huang, K. Muraoka, Y. Wang, J. Chang, F. Yuan, S.L. Bore, C. Cai, Y. Lin, B. Wang, J. Xu, J.-X. Zhu, C. Luo, Y. Zhang, R.E.A. Goodall, W. Liang, A.K. Singh, S. Yao, J. Zhang, R. Wentzcovitch, J. Han, J. Liu, W. Jia, D.M. York, W. E, R. Car, L. Zhang, H. Wang, DeePMD-kit v2: A software package for deep potential models, The Journal of Chemical Physics 159 (2023) 054801.

[S3] M.J. Verstraete, J. Abreu, G.E. Allemand, B. Amadon, G. Antonius, M. Azizi, L. Baguet, C. Barat, L. Bastogne, R. Béjaud, J.-M. Beuken, J. Bieder, A. Blanchet, F. Bottin, J. Bouchet, J. Bouquiaux, E. Bousquet, J. Boust, F. Brieuc, V. Brousseau-Couture, N. Brouwer, F. Bruneval, A. Castellano, E. Castiel, J.-B. Charraud, J. Clérouin, M. Côté, C. Duval, A. Gallo, F. Gendron, G. Geneste, P. Ghosez, M. Giantomassi, O. Gingras, F. Gómez-Ortiz, X. Gonze, F.A. Goudreault, A. Grüneis, R. Gupta, B. Guster, D.R. Hamann, X. He, O. Hellman, N. Holzwarth, F. Jollet, P. Kestener, I.-M. Lygatsika, O. Nadeau, L. MacEnulty, E. Marazzi, M. Mignolet, D.D. O'Regan, R. Outerovitch, C. Paillard, G. Petretto, S. Poncé, F. Ricci, G.-M. Rignanese, M. Rodriguez-Mayorga, A.H. Romero, S. Rostami, M. Royo, M. Sarraute, A. Sasani, F. Soubiran, M. Stengel, C. Tantardini, M. Torrent, V. Trinquet, V. Vasilchenko, D. Waroquiers, A. Zabalo, A. Zadoks, H. Zhang, J. Zwanziger, Abinit 2025: New capabilities for the predictive modeling of solids and nanomaterials, The Journal of Chemical Physics 163 (2025) 164126.

[S4] J. P. Perdew, A. Ruzsinszky, G. I. Csonka, O. A. Vydrov, G. E. Scuseria, L. A. Constantin, X. Zhou, and K. Burke, Restoring the density-gradient expansion for exchange in solids and surfaces, Physical Review Letters 100, (2008) 136406.

[S5] A. Hjorth Larsen, J. Jørgen Mortensen, J. Blomqvist, I.E. Castelli, R. Christensen, M. Dułak, J. Friis, M.N. Groves, B. Hammer, C. Hargus, E.D. Hermes, P.C. Jennings, P. Bjerre Jensen, J. Kermode, J.R. Kitchin, E. Leonhard Kolsbjerg, J. Kubal, K. Kaasbjerg, S. Lysgaard, J. Bergmann Maronsson, T. Maxson, T. Olsen, L. Pastewka, A. Peterson, C. Rostgaard, J.

Schiøtz, O. Schütt, M. Strange, K.S. Thygesen, T. Vegge, L. Vilhelmsen, M. Walter, Z. Zeng, K.W. Jacobsen, The atomic simulation environment - a Python library for working with atoms, Journal of Physics: Condensed Matter 29 (2017) 273002.

[S6] A.P. Thompson, H.M. Aktulga, R. Berger, D.S. Bolintineanu, W.M. Brown, P.S. Crozier, P.J. in 't Veld, A. Kohlmeyer, S.G. Moore, T.D. Nguyen, R. Shan, M.J. Stevens, J. Tranchida, C. Trott, S.J. Plimpton, LAMMPS - a flexible simulation tool for particle-based materials modeling at the atomic, meso, and continuum scales, Computer Physics Communications 271 (2022) 108171.

[S7] L. Mao, Q. Meng, A. Ahmad, Z. Wei, Mechanical analyses and structural design requirements for flexible energy storage devices. Advanced Energy Materials 7 (2017) 1700535.

[S8] M. Rashid, Q. Mahmood, F. Babar, S.M. Ramay, A. Mahmood, Study of mechanical, electronic and optical properties of $PbZrO_3$ and $PbHfO_3$; DFT approach. Materials Research Express 6 (2019) 066311.

[S9] P. Vales-Castro, K. Roleder, L. Zhao, J.-F. Li, D. Kajewski, G. Catalan, Flexoelectricity in antiferroelectrics. Applied Physics Letters 113 (2018) 132903.